\documentclass[conference]{IEEEtran}

\usepackage{graphicx}
\usepackage{amsmath,url}
\usepackage{multirow}
\usepackage{hhline}
\usepackage{rotating}
\usepackage{color, colortbl}
\definecolor{Gray}{gray}{0.9}
\usepackage{amssymb}
\usepackage{array}
\usepackage{tabularx}
\usepackage{fancyhdr}
\newcolumntype{L}[1]{>{\raggedright\let\newline\\\arraybackslash\hspace{0pt}}m{#1}}
\newcolumntype{C}[1]{>{\centering\let\newline\\\arraybackslash\hspace{0pt}}m{#1}}
\newcolumntype{R}[1]{>{\raggedleft\let\newline\\\arraybackslash\hspace{0pt}}m{#1}}
\usepackage{color,soul}

\ifCLASSINFOpdf

\fi

\begin{document}

\title{Novel Quality Metric for Duration Variability Compensation in Speaker Verification using i-Vectors}

\author{\IEEEauthorblockN{Arnab Poddar\IEEEauthorrefmark{1},
Md Sahidullah\IEEEauthorrefmark{2},
Goutam Saha\IEEEauthorrefmark{4}}
\vspace{0.35cm}

\IEEEauthorblockA{\IEEEauthorrefmark{1}\IEEEauthorrefmark{4} Dept of Electronics and Electrical Communication Engineering,
Indian Institute of Technology, Kharagpur, India
}

\IEEEauthorblockA{\IEEEauthorrefmark{2}Speech and Image Processing Unit, School of Computing, University of Eastern Finland, Joensuu, Finland}
\vspace{0.15cm}
\IEEEauthorblockA{Email: \IEEEauthorrefmark{1}arnabpoddar@iitkgp.ac.in, \IEEEauthorrefmark{2}sahid@cs.uef.fi, \IEEEauthorrefmark{4}gsaha@ece.iitkgp.ernet.in}}

\maketitle

\begin{abstract}
Automatic speaker verification (ASV) is the process to recognize persons using voice as biometric. The ASV systems show considerable recognition performance with sufficient amount of speech from matched condition.
One of the crucial challenges of ASV technology is to improve recognition performance with speech segments of short duration. In short duration condition, the model parameters are not properly estimated due to inadequate speech information, and this results poor recognition accuracy even with the state-of-the-art i-vector based ASV system. We hypothesize that considering the estimation quality during recognition process would help to improve the ASV performance. This can be incorporated as a quality measure during fusion of ASV systems. This paper investigates a new quality measure for i-vector representation of speech utterances computed directly from Baum-Welch statistics. The proposed metric is subsequently used as quality measure during fusion of ASV systems. In experiments with the NIST SRE 2008 corpus, We have shown that inclusion of proposed quality metric exhibits considerable improvement in speaker verification performance. The results also indicate the potentiality of the proposed method in real-world scenario with short test utterances.

\end{abstract}

\thispagestyle{fancy}

\fancyhf{}

\renewcommand{\headrulewidth}{0pt}

\chead{\small Published in Ninth International Conference on Advances in Pattern Recognition (ICAPR-2017), Bangalore, India}

\pagestyle{empty}

\lfoot{\small \copyright 2017 IEEE. Personal use of this material is permitted. Permission from IEEE must be obtained for all other uses, in any current or future media, including reprinting/republishing this material for advertising or promotional purposes, creating new collective works, for resale or redistribution to servers or lists, or reuse of any copyrighted component of this work in other works.}


\IEEEpeerreviewmaketitle

\section{Introduction}
  Automatic speaker verification (ASV) is a biometric recognition system where the voice is used as the trait \cite{kinnunen2010overview,reynolds2000speaker}. ASV is a convenient and non-invasive technology that can potentially be applied to various important applications, covering access of control, authentication of secure transactions
over a telephone connection and forensic identification of suspects using
voice samples \cite{kinnunen2010overview,poddar2017speaker}. Contrasting to other biometrics, speaker recognition is a non-obtrusive technology and does not involve special purpose acquisition hardware other than a microphone. Even though speaker recognition research has been ongoing for more than four decades, the state-of-the-art speaker recognition systems still have several limitations\cite{poddar2017speaker,arnab2015comparison,li2016improving}.

  Although state-of-the-art i-vector based  ASV systems exhibit satisfactory performance with adequate speech data, but practically, the performance of such systems decline with limited duration data  \cite{kanagasundaram2011vector,poddar2017speaker,arnab2015comparison}. ASV system, in real-world applications requires satisfactory performance with short duration speech which remains as an opportunity to explore further. The work in \cite{hasan2013duration} attempted to model the duration variability in short duration as noise and also compensated with synthetically generated supporting i-vectors for speaker modeling. The work in \cite{kanagasundaram2014improving} proposed to estimate the variability originated due to shorter utterances in i-vector space. The ASV systems suffers from the duration variability due to mismatch in train-test segments.
\par
 In the modern i-vector based ASV systems, Baum-Welch (BW) statistics are indispensable intermediate parameters, which totally represent the extracted speech-features. The quality of estimation of BW statistics is degraded in short duration, which introduces sparsity due to insufficient data. The sparsity arises due to shortage of speech data as it fails to update most of the Gaussian components in BW statistics \cite{arnab2017adapt,li2016feature}. The present de-facto ASV systems do not include the information regarding the quality of speaker model estimation. We consider BW statistics not only as the intermediate  parameters for speaker model estimation, but also as a source to determine quality of speaker model estimation.

\begin{figure*}[t]
\centering
\includegraphics[width=14cm]{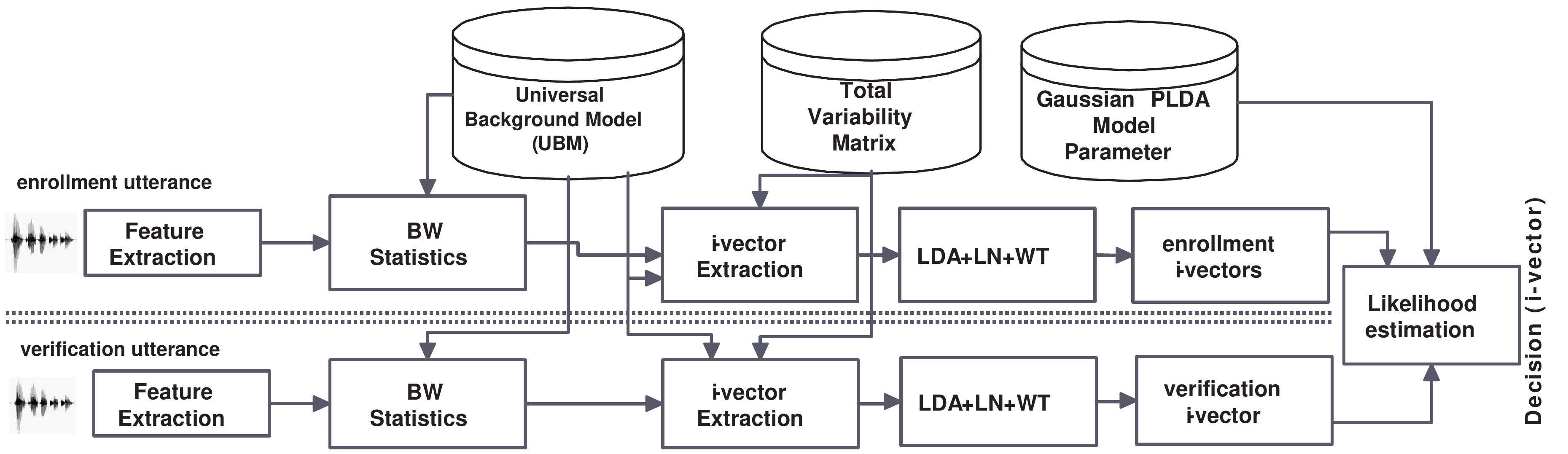}
\caption{Block diagram for i-vector based ASV system.}
\label{ivsys}
\end{figure*}

 \par 
 In this work, we introduce a metric to measure the quality of intermediate ASV system parameters. This work contributes to incorporate the information regarding quality of the speaker model estimation for the first time in ASV to the best of our knowledge. the proposed metric is estimated directly from the intermediate system parameters of i-vector based ASV system. This metric attempt to represent the impact of duration on intermediate BW statistics by calculating the difference between intermediate BW statistics and universal background parameters.
 The proposed dissimilarity metric do not require additional parameters to be estimated and  require negligible computation cost as intermediate statistics are inherently calculated by state-of-the-art ASV systems.
    In the classification module of ASV systems, \emph{Gaussian mixture model-universal background model} (GMM-UBM) \cite{reynolds2000speaker}  and i-vector \cite{dehak2011front} are used widely.  An exhaustive comparison of the two techniques, including the short duration effect, reveal that though i-vector  outperforms the GMM-UBM for longer speech utterances, but the GMM-UBM is considerably relevant for short duration condition~\cite{arnab2015comparison}. The observation inspire us to fuse classifiers. Additionally, the proposed similarity metric is incorporated in fusion stage as quality information of speech to compensate the short duration effect. Incorporation of quality measures not only showed considerable improvement in performance in various duration conditions. The proposed systems showed more improvement for practical requirement i.e., in short duration cases.

\begin{figure}[t]
\centering
 \includegraphics[height=5.5cm]{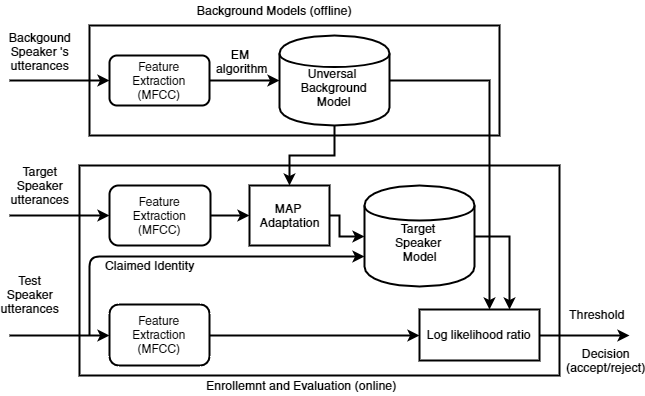}
\caption{Block diagram for GMM-UBM based ASV system.}
\label{ubmsys}
\vspace{-0.5cm}
\end{figure}

In the rest of the paper, theoretical aspects of widely used i-vector and GMM-UBM based ASV techniques are illustrated in Section \ref{Theory}.
An analysis on intermediate parameters is presented in section \ref{SUC}. Subsequently, Section \ref{ProposedQuality} and \ref{Exp_set} discuss the proposed quality aided fusion based system and experimental results. Finally, we conclude the paper  in Section \ref{conclude}.
\vspace{-0.25cm}

\section{ASV Systems}\label{Theory}
Here, we discuss two popularly implemented ASV techniques, namely GMM-UBM~\cite{reynolds2000speaker} and  i-vector representation of speech utterance~\cite{dehak2011front}. Fig.~\ref{ivsys} shows i-vector based the ASV system and Fig.~\ref{ubmsys} represents the GMM-UBM system.

\subsection{GMM-UBM system}\label{section1}
In GMM-UBM approach, initially, a GMM is estimated with a large volume of voice from a large number of speakers who may not participate in the verification process~\cite{reynolds2000speaker}. The estimated background model is termed as \emph{universal background model} (UBM), written as $\mathbf{\lambda}_{\mathrm{UBM}}=\{ w_{i},\bar{\mathbf{\mu}}_{i},\bar{\mathbf{\Sigma}}_{i};\space i=1,2,\dots,K\}$. Here, $K$ represents the number of Gaussian components in the mixture, $w_{i}$ stands for the prior weights of the $i$-th Gaussian mixture components, $ \mathbf{\mu}_{i}$ represents the mean and  $\mathbf{\Sigma}_{i}$ represents the co-variance matrix. The parameter $w_{i}$ meets the condition $\sum_{i=1}^{K}w_{i}=1$.
\par
The $S$ speakers' GMM models are mathematically represented as $ \{ \mathbf{\lambda}_{1},\mathbf{\lambda}_{2},\ldots, \mathbf{\lambda}_{S} \} $. We estimate the enrollment speaker model by adapting of the UBM model parameters using \emph{maximum-a-posteriori} (MAP)  technique~\cite{reynolds2000speaker}. Initially, sufficient statistics $N_{i}$ (zero order),  $\mathbf{E}_{i}$ (1st order)  and  $\mathbf{F}_{i}$ (2nd order) from a enrollment speaker's utterance with $C$ active frames $\mathbf{X} =  \{ \mathbf{x}_{1}, \mathbf{x}_{2},\ldots, \mathbf{x}_{C} \}$, are computed as,
\begin{equation}
\label{eq:ni}
  N_{i}= \sum_{t=1}^{C}Pr(i|\mathbf{x}_{t}),
\end{equation}
\begin{equation}
\mathbf{E}_{i}(\mathbf{X})= \frac{1}{N_{i}} \sum_{t=1}^{C}Pr(i|\mathbf{x}_{t})\mathbf{x}_{t},
\end{equation}
Here the distribution of probability of Gaussian mixture components $ Pr(i|\mathbf{x}_{t}) $ for given speech segments with $C$ frames $\mathbf{X}^{\mathrm{train}} =  \{ \mathbf{x}_{1}, \mathbf{x}_{2},\ldots, \mathbf{x}_{C} \}$
 is formulated by
  \begin{equation}
   Pr(i|\mathbf{x}_{t})= \frac{w_{i}p_{i}(\mathbf{x}_{t})}{\sum_{j=1}^{K}w_{j}p_{j}(\mathbf{x}_{t})}
  \end{equation}
where all probability density is a $K$-dimensional Gaussian variable of the form
\begin{equation}
\label{GaussDist}
p_{i}(\mathbf{x})=\frac{1}{(2\pi)^{K/2}|\bar{\mathbf{\Sigma}}|^{\frac{1}{2}}}exp \{ -\frac{1}{2}(\mathbf{x}- \bar{\mathbf{\mu}}_{i})^{\top} \bar{\mathbf{\Sigma}}_{i}^{-1} (\mathbf{x}- \bar{\mathbf{\mu}}_{i}) \}.
\end{equation}
Conventionally,  only the mean parameters of speaker's GMM model are adapted to estimate the enrollment models. It makes the speaker model estimation process computationally efficient ~\cite{reynolds2000speaker}.
\par
During evaluation, the log-likelihood ratio of verification feature vectors are computed, $ \mathbf{X}^{\mathrm{test}}= \{\mathbf{x}_{1}, \mathbf{x}_{2}, \ldots, \mathbf{x}_{C} \} $ against both target-speaker model and the background model as,
\begin{equation}
\Lambda_{\mathrm{GMM-UBM}}(\mathbf{X}^{\mathrm{test}})= \log\ p(\mathbf{X}^{\mathrm{test}}|\mathbf{\lambda}_{\mathrm{target}})- \log\ p(\mathbf{X}^{\mathrm{test}}|\mathbf{\lambda}_{\mathrm{UBM}})
\end{equation}
Finally, a threshold ($\theta$) is adjusted to determine whether the claimed identity will be \emph{accepted} or \emph{rejected}.  If $\Lambda_{\mathrm{GMM-UBM}}(\mathbf{X}) \geq \theta$, the claim is accepted, otherwise rejected.


\subsection{i-vector Extraction}\label{TVs}
The i-vectors transform the \emph{GMM supervector} into a lower dimensional subspace~\cite{dehak2011front}.  The adapted GMM supervector of $i-th$ speaker, $\mathbf{m}_{i}$, is transformed as,
\begin{equation}
  \mathbf{m}_{i}=\bar{\mathbf{m}}+\mathbf{\Phi y},
\end{equation}
here $\mathbf{\Phi}$ is a matrix of lower-rank, denoting the  channel and speaker independent subspace, $\mathbf{y}$ is i-vector, $\bar{\mathbf{m}}$ represents the channel and speaker independent supervector ($\bar{\mathbf{m}}$). Initially, $\mathbf{\Phi}$ is estimated with large volume of voice utterances collected from various persons \cite{dehak2011front}. Subsequently the corresponding i-vectors are calculated with the $1^{st}$ order and zeroth order BW statistics $\mathbf{E}_i$ and $N_i$, respectively.

Initially, sufficient statistics $N_{i}$ (zero order),  $\mathbf{E}_{i}$ ($1^{st}$ order)  and  $\mathbf{F}_{i}$ ($2^{nd}$ order) from a speaker's voice segment, consisting of $C$ active frames $\mathbf{X} =  \{ \mathbf{x}_{1}, \mathbf{x}_{2},\ldots, \mathbf{x}_{C} \}$, are computed as,
\begin{equation}
\label{eq:ni}
  N_{i}= \sum_{t=1}^{C}Pr(i|\mathbf{x}_{t}),
\end{equation}
\begin{equation}
\mathbf{E}_{i}(\mathbf{X})= \frac{1}{N_{i}} \sum_{t=1}^{C}Pr(i|\mathbf{x}_{t})\mathbf{x}_{t},
\end{equation}
here the distribution of probability of Gaussian components $ Pr(i|\mathbf{x}_{t}) $ conditioned on given voice segment with $C$ speech frames $\mathbf{X}^{\mathrm{train}} =  \{ \mathbf{x}_{1}, \mathbf{x}_{2},\ldots, \mathbf{x}_{C} \}$
 is given by
  \begin{equation}
   Pr(i|\mathbf{x}_{t})= \frac{w_{i}p_{i}(\mathbf{x}_{t})}{\sum_{j=1}^{K}w_{j}p_{j}(\mathbf{x}_{t})}
  \end{equation}
where each component density is a $d$-variate Gaussian function of the form as shown in Eq. \ref{GaussDist}.
We consider the prior distribution of i-vectors $p(\mathbf{y})$ is normally distributed as $ \mathcal{N}(0,\mathbf{I})$. The corresponding posterior distribution of $p(\mathbf{E|y})$, is assumed as $p(\mathbf{E|y})=\mathcal{N}(\mathbf{\Phi} \mathbf{y}, \mathbf{N}^{-1}\mathbf{\Sigma})$. The intermediate parameter $\mathbf{N}$ is computed as a diagonal matrix having $\mathbf{N}_i$ as its diagonal elements~\cite{dehak2011front}. The MAP estimate of ($\mathbf{y|E}$) is computed as

\begin{equation}
\label{iveceq}
  \mathbb{E}(\mathbf{y|E})=(\mathbf{I}+ \mathbf{\Phi}^{\top} \mathbf{\Sigma}^{-1}\mathbf{N\Phi)^{-1}\Phi}^{\top}\mathbf{\Sigma}^{-1}\mathbf{N}(\mathbf{E}-\bar{\mathbf{m}})
\end{equation}
The expectation of ($\mathbf{y|E}$) is termed as the i-vector of a given voice segment $\mathbf{X}$ ~\cite{dehak2011front}.



The verification scores in i-vector GPLDA framework, is calculated as the likelihood ratio~\cite{kenny2010bayesian}. For a verification trial, the projected verification and enrollment i-vectors $\mathbf{z}_{test}$ and $\mathbf{z}_{target}$ respectively are used to estimate the  likelihood ratio $\Lambda_{\mathrm{GPLDA}}( \mathbf{z}_{\mathrm{target}}, \mathbf{z}_{\mathrm{test}} )$ as,
\begin{equation}
  \Lambda_{\mathrm{GPLDA}}(\mathbf{z}_{\mathrm{target}}, \mathbf{z}_{\mathrm{test}})=log\ \frac{ \space p(\mathbf{z}_{\mathrm{target}}, \mathbf{z}_{\mathrm{test}}|H_1)} { p(\mathbf{z}_{\mathrm{target}}|H_0)\ p(\mathbf{z}_{\mathrm{test}}|H_0)}
\end{equation}
here $H_1$ hypothesizes the projected i-vectors belong to the same person. On the other hand, $H_0$ denotes the hypothesis where the i-vectors belong to different person.

\section{Analysis of and Characteristics of BW statistics}
\label{SUC}
BW statistics represent the overall extracted information from the speech and are transformed into i-vectors using the pre-estimated \emph{universal background model} (UBM) \cite{dehak2011front,reynolds2000speaker}.
\begin{figure}[t]
\centering
 \includegraphics[height=8cm]{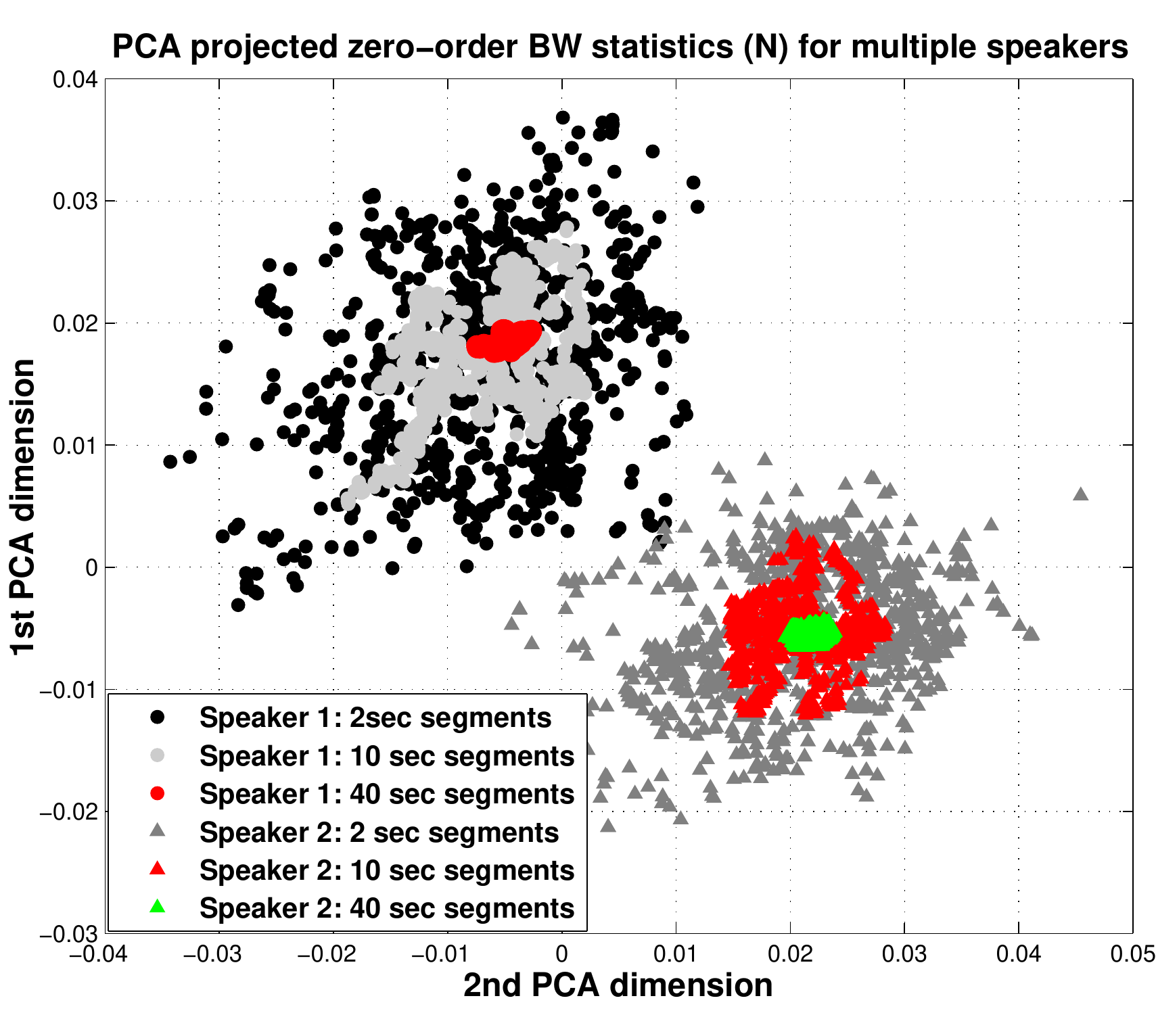}
\caption{Scatter plot of PCA projected NBS ($\tilde{N}$) for two speakers.}
\label{NScatterTruncated}
\vspace{-0.5cm}
\end{figure}

\begin{figure*}
\centerline{
\includegraphics[width=17cm]{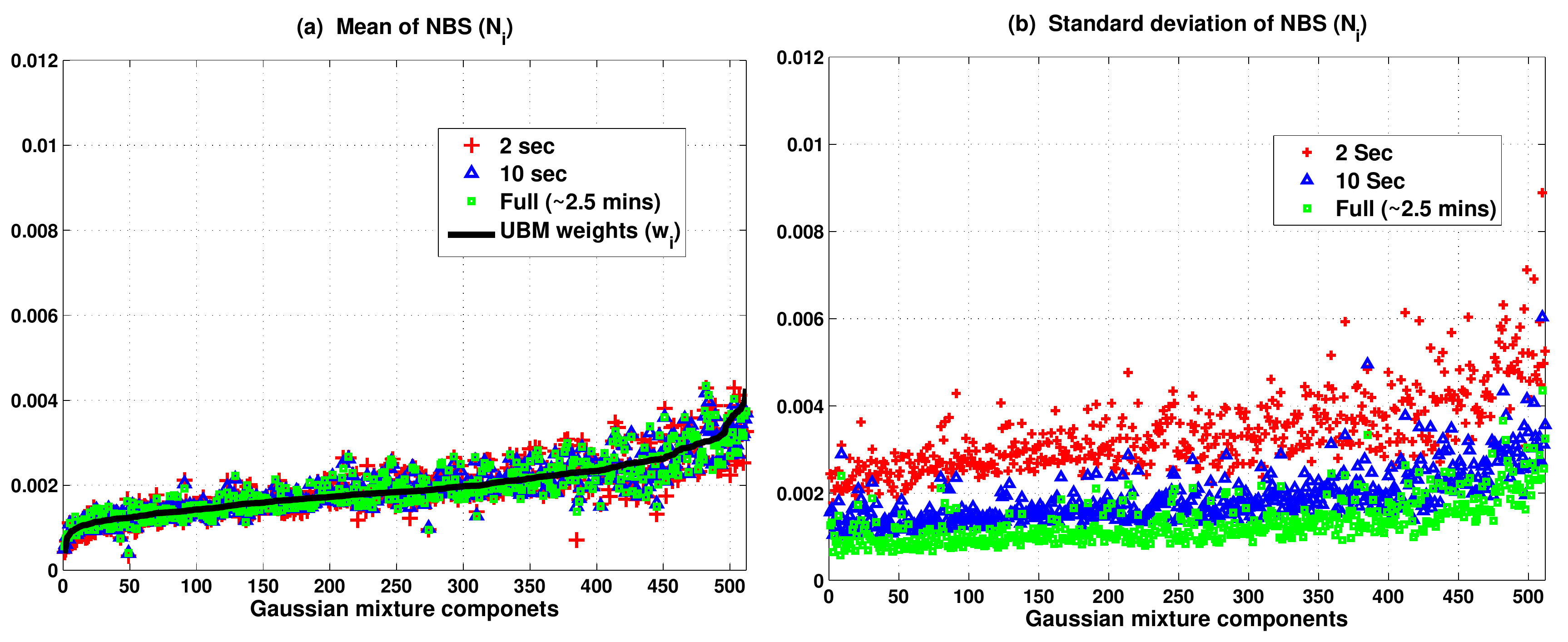}}

 \caption{Mean and standard deviation of NBS, calculated from voice utterances of 1270 male speakers (NIST 2008), are presented in (a) and (b) respectively.  Means and standard deviation of NBS are plotted for three duration conditions, e.g., 2 Sec, 10 Sec and full length (~2.5 min). Weights of GMM-UBM for corresponding Gaussian mixture component are shown in (a).}
\label{NdistUBM}
\vspace{-0.5cm}
\end{figure*}

Since BW statistics is an indispensable intermediate step in ASV, we investigate its characteristics in short duration. The zeroth order BW statistics ($N_i$) is estimated as,
$
N_i= \displaystyle \sum_{t=1}^{C}Pr(i|\mathbf{x}_{t})
$
, where, a speech segment with $C$ frames is represented as $\mathbf{X} =  \{ \mathbf{x}_{1}, \mathbf{x}_{2},\ldots, \mathbf{x}_{K} \}$, and $Pr()$ is the prior probability of i-th Gaussian component. Summing over all Gaussian mixture components $K$ we obtain,
\begin{equation}
\sum_{i=1}^{K}N_{i}= \displaystyle \sum_{i=1}^{K}\sum_{t=1}^{C}Pr(i|\mathbf{x}_{t}) = C.
\end{equation}
The equation indicates that $N_i$ is dependent on segment duration, i.e, $C$.
Normalizing $N_i$ with number of frames ($C)$ we get

\begin{equation}
\label{ni}
\tilde{N}_i= \displaystyle \frac{1}{C}\sum_{t=1}^{C}Pr(i|\mathbf{x}_{t})
\end{equation}
and
$
\sum_{i=1}^{C}\tilde{N}_i=1
$
 hence it has the same property as weights of the GMM UBM i.e., $\sum_{i=1}^{C}w_i=1 $ .

 Now, $\tilde{N}_i$ can be regarded as the mixture weights of the Gaussian components $i$ for a particular speech segment $s$.
  It is a standard statistical hypothesis that intermediate BW statistics can be estimated more efficiently with sufficient volume speech corpus, which is likely to include all possible kinds of variability proportionately. Hence, we expect that  large number of speech frames ($C$) in the speech segment would be advantageous for improvement of quality of $\tilde{N}_i$. However, the intermediate statistics may be expected to be updated more sparsely due to reduced speech data or degraded quality of speech.   On this core note, the characteristics of $\tilde{N}_i$ are investigated.

 Here we present analysis on NBS and its characteristics for different duration conditions, to observe the impact of duration. We considered the recordings from telephonic conversations in NIST SRE 2008 (\emph{short2}) corpus. Truncated voice segments  (40, 10, 2 sec) are considered for the analytical experiments. For  truncation of the long segments, the initial speech frame is  selected randomly and required number of successive active frames are pruned. Similarly, 1000 truncated utterances are generated  for the duration conditions, under consideration.

 We apply the principal component analysis (PCA) on the feature matrix. Subsequently, we show the major two projected components. In Fig.\ref{NScatterTruncated}, the projected components of different truncated segments of 2 Sec, 10 Sec and 40 Sec are shown. We estimate the matrix for PCA projection from the generated 1000 truncated segments.
We observe that the NBS show greater variability in limited duration. Higher variability in NBS for limited duration, deteriorates the quality in i-vector model. The change in variability of BW statistics with duration of the speech segments indicate that BW statistics is associated with duration or estimation quality.

Fig. \ref{NdistUBM} (a) and (b) represents the mean and standard deviation of NBS ($\tilde{N}_i$) respectively, for three duration conditions (2 sec, 10 sec and full length). The means of NBS ($\tilde{N}_i$) are calculated using 1270 male speakers from NIST 2008 telephone corpus. The weights of GMM-UBM of corresponding mixture components $(w_i)$ are presented simultaneously in Fig. \ref{NdistUBM} (a).
The short segments in Fig. \ref{NdistUBM} (b) showed greater standard deviation referring greater variability introduced in NBS ($\tilde{N}_i$). We observe gradual increment in variability when the length of speech segments are shortened. As the variability in NBS is affected by duration, we hypothesize that the information of duration variability can be estimated from NBS and also can be treated as the source of the information about the quality of speaker-model estimation.
It is observed in  Fig. \ref{NdistUBM} (a) that the means of $\tilde{N}$ for different duration condition follows the value of UBM weight of corresponding Gaussian mixture component $(w_i)$. We also observe that the means of different duration condition for a particular Gaussian component remains nearly equal. This is observed in almost all Gaussian components shown in Fig. \ref{NdistUBM} (a). These observations on means of $\tilde{N}$ distribution and weights of corresponding Gaussian mixture component inspired us to use GMM-UBM weights $(w_i)$ as reference to measure the variability in $\tilde{N}$.







\begin{figure*}[t]
\centering
 \includegraphics[height=4cm]{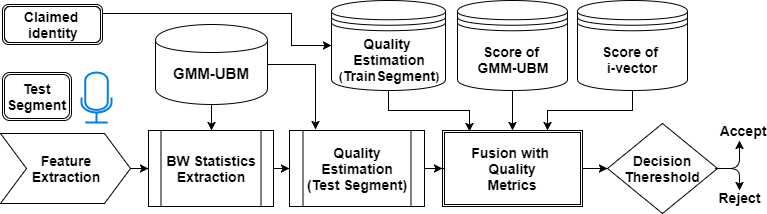}
\caption{Block diagram for the proposed quality estimation and quality metric incorporated  fusion based approach for ASV.}
\label{qual}
\end{figure*}

\section{Proposed Quality Measure and ASV system Fusion}
\label{ProposedQuality}
The observations in previous section demonstrate that the sparsity  in BW statistics is associated with duration of speech and quality of speaker model estimation. The sparsity in $\tilde{N}$ increases in short duration, indicating lower quality of estimation. From the observations in Section \ref{SUC}, we consider that the BW
statistics not only as intermediate parameters but also the source of estimating the quality of speaker model estimation.
Here, we propose to quantify the dissimilarity between normalized zeroth order BW statistics $\tilde{N}_i$ and prior of corresponding Gaussian component of UBM model $w_i$. We further use it as a quality metric. Subsequently, it is incorporated as supporting information in proposed ASV system.
The mathematical expressions to model the quality $Q$ of a segment $s$ is given by
\begin{equation}
Q_s(\tilde{N}_s)=\sum^{K}_{i=1}|\tilde{N}_{i,s}-w_{i,ubm}|
\end{equation}
The proposed quality metric attempts to measure dissimilarity of $\tilde{N}_i$  from the weights of UBM $(w_i)$ which is treated as reference.

\begin{table}[t]
  \centering

  \caption{Details of speech corpus and cepstral features.}

  \begin{tabular}{| l | c | c|c|c|}
  \hline
    \textbf{Specifications} & \#target & \#test & \#genuine &\#imposter   \\
 &  model& segments&  trials& trials   \\
    \hline
 \textbf{NIST 2008} & 442 & 854 & 874 & 11637\\
  \hline\hline
  \multicolumn{5}{|c|}{\cellcolor{Gray}{\textbf{specifications: Features and Development parameters}}}\\ \hline \hline
  \textbf{MFCC}  & \multicolumn{4}{l|}{Dimension: 19+19$\Delta$+19$\Delta\Delta$; 20ms Hamming}\\\hline
 \textbf{GMM-UBM} & \multicolumn{4}{l|}{Dimension: 512}\\
  & \multicolumn{4}{l|}{Data: NIST SRE '04, '05, Switchboard II}\\\hline
  \textbf{TV ($\Phi$) Matrix}  & \multicolumn{4}{l|}{Dimension: 400;}\\
    & \multicolumn{4}{l|}{Data: NIST SRE  '04, '05, '06, Switchboard II}\\\hline
  \textbf{GPLDA} & \multicolumn{4}{l|}{Dimension: 150;}\\
  & \multicolumn{4}{l|}{Data: NIST SRE  '04, '05, '06, Switchboard II}\\
   \hline
  \end{tabular}

  \label{DatabaseSummary}

\end{table}

ASV systems based on fusion approach have found wide applications   \cite{hasan2013crss,hautamaki2013sparse}. Though i-vector  \cite{dehak2011front,kanagasundaram2014improving} and GMM-UBM \cite{reynolds2000speaker} based ASV systems have different modeling approaches, they exhibit similar performance in short utterance cases \cite{arnab2015comparison}.
Here, we  exploit the information captured simultaneously by GMM-UBM and modern i-vector GPLDA using linear fusion and subsequently incorporate the quality metric $Q$ as additional information. The fusion parameters are trained using logistic regression objective using the BOSARIS toolkit \cite{brummer2013bosaris}.
 We confine our work to score level fusion with fusion function $f$ which combines two base classifier score $\Lambda_{\mathrm{UBM}}$ and $\Lambda_{\mathrm{i-vector}}$ into a single match score $\mathbf{\Lambda}=\{\Lambda_{\mathrm{UBM}},\Lambda_{\mathrm{i-vector}}\}^{\top}$. The decision is made by a predefined score threshold $\theta$. The trained linear fusion classifier is of the form
\begin{equation}
f_{\mathbf{\alpha},\theta}(\mathbf{\Lambda})= \mathbf{\alpha}^{\top}\mathbf{\Lambda}+\theta
\end{equation}
The fusion function, incorporating quality measure $Q$ is represented by:
\begin{equation}
  f_{Q}(\mathbf{\Lambda})= \mathbf{\alpha}^{\top}\mathbf{\Lambda}+\theta + \beta *Q(\tilde{N}_{\mathrm{enrollment}}) Q(\tilde{N}_{\mathrm{verification}})
\end{equation}
 A speaker is accepted if and only if $f_{Q, \mathbf{\alpha}, \beta, \theta}(\mathbf{\Lambda})\geq 0$. When for a quality fusion classifier $f_Q(\mathbf{\Lambda})$ with parameters $(\mathbf{\alpha},\beta,\theta)$, the development data $D$ and an empirical cost function $\hat C((\mathbf{\alpha},\beta,\theta),D)$ are given, the optimal fusion device is obtained by $(\mathbf{\alpha}^{dev},\beta^{dev},\theta^{dev})= arg min_{\mathbf{\alpha},\beta,\theta}\hat C((\mathbf{\alpha},\beta,\theta),D)$. Here, the \emph{decision cost function} is adopted as
\begin{equation}
  C_{\mathrm{det}}(\theta)=C_{\mathrm{miss}}P_{\mathrm{miss}}(\theta)P_{\mathrm{tar}}+C_{\mathrm{fa}}P_{\mathrm{fa}}(\theta)(1-P_{\mathrm{tar}})
\end{equation}

where $P_{tar}$ is the prior probability of an original speaker, $C_{miss}$ is the cost of a miss and $C_{fa}$ is the cost of false alarm.
A diagrammatic representation of the proposed fusion based approach to include the quality metrics in presented in Fig.~\ref{qual}.

\begin{table}[t]
 \centering

\caption{Results of fusion of GMM-UBM and i-vector based system with proposed quality metric on NIST 2008 \emph{Truncated Train - Truncated Test} telephone corpora}
 \setlength{\tabcolsep}{5pt}
\label{TTabDevEER}  

\begin{tabular}{|l|c||cccc|}
\noalign{\smallskip}\hline

Train-Test &Metric&GMM& i-vect & linear & Quality  \\

duration&&UBM& GPLDA & fusion & fusion \\
\hline\hline

2s-2s &EER & 35.24 & 36.84 & 33.05 & \textbf{31.92}   \\
 &DCF& 9.69& 9.93 & 9.54  & \textbf{9.50} \\
5s-5s &EER& 25.25& 24.37 & 23.11 & \textbf{21.25} \\
& DCF& 8.89&8.65 & 8.27  & \textbf{8.13} \\
10s-10s &EER& 14.98 & 14.53 & 14.05 & \textbf{13.15} \\
 &DCF&  6.58 & 6.40 & 6.25 & \textbf{5.86}\\
\hline
\end{tabular}

\end{table}

\begin{table}[t]
 \centering

\caption{Results of fusion of GMM-UBM and i-vector based system with proposed quality metric on NIST 2008 \emph{Long Train- Truncated Test} telephone corpora}
 \setlength{\tabcolsep}{5pt}
\label{TFTabDevEER}  

\begin{tabular}{|l|c||cccc|}
\noalign{\smallskip}\hline

Train-Test &Metric&GMM& i-vect & linear & Quality  \\

duration&&UBM& GPLDA & fusion & fusion \\
\hline\hline
Full-2s & EER&21.56& 19.67 & 19.10 &  \textbf{16.81}    \\
& DCF& 7.75 &7.91& 7.53 & \textbf{7.01}  \\
Full-5s &EER&17.73& 13.50 & 12.23 & \textbf{11.67}  \\
 &DCF& 7.32&5.99 & 5.69 & \textbf{5.42} \\
Full-10s &EER&16.66& 9.29 & 9.72 & \textbf{9.09}   \\
 &DCF&6.75 &4.50 & 4.59  & \textbf{4.28} \\
\hline
\end{tabular}

\end{table}

\section{Experimental Results and Discussion}\label{Exp_set}

  Here, Mel-frequency cepstral coefficients (MFCC) appending delta ($\Delta$) and double-delta ($\Delta\Delta$) coefficients are used for experiments \cite{sahidullah2016local}. The non-speech frames are discarded by an spectrum energy-based detector (SAD) and finally, cepstral mean and variance normalization (CMVN) is applied as feature normalization~\cite{sahidullah2012design,sahidullah2013novel}.  A gender-specific UBM (male) is used. We conducted the experiments  on NIST speaker recognition evaluation (SRE) corpus 2008. We considered \emph{short2-short3} task\footnote{\url{http://www.itl.nist.gov/iad/mig/tests/sre/2008/sre08_evalplan_release4.pdf}}
 on \emph{telephone-telephone} part of male speakers. The details of NIST SE 2008 are given in Table \ref{DatabaseSummary}. Channel compensation for i-vectors are done using Gaussian probabilistic linear discriminant analysis (GPLDA) \cite{kenny2010bayesian,kanagasundaram2011vector,arnab2015comparison}. A brief synopsis of the development parameters used in the experiments are outlined in Table~\ref{DatabaseSummary}.  To generate short utterances, we truncate the long speech utterances in 2 sec (200 active frames), 5 sec (500 active frames), 10 sec (1000 active frames) duration removing prior 500 active speech frames to avoid phonetic similarity in initial greetings to avoid text-dependence \cite{arnab2017adapt,arnab2017improved}.





Quality measures of speech signals are used to support the fusion based ASV system. Performance measures of GMM-UBM, i-vector and fusion, using quality measures are depicted separately in Table \ref{TTabDevEER} and Table \ref{TFTabDevEER}. Separate experiments are conducted with long enrollment data (\ref{TFTabDevEER}) and also short enrollment data (\ref{TTabDevEER}). A total of six different duration conditions are used for experiments as shown in Table \ref{TTabDevEER} and \ref{TFTabDevEER}. The results are shown in both \emph{equal error rate} (EER) and \emph{minimum detection cost function} (minDCF)~\cite{kanagasundaram2011vector}. Incorporation of quality metric exhibited considerably high relative improvement over state-of-the-art, in conditions like \emph{full-2 sec, full-5 sec, 5 sec-5 sec, 2 sec-2 sec} etc. These conditions are more close to desirable real-time requirements of ASV systems which encourages to find implementations of proposed system. Consistent improvement of accuracy of the ASV system in various duration  established relevance of the proposed quality measures based on intermediate statistics. The system is more suitable when the duration of speech utterances are limited, especially when it is trained with long enrollment data and tested with very short duration of speech.

\section{Conclusion}
\label{conclude}
This work investigates a new metric for measuring the quality of the i-vector estimation process. The metric is formulated using the Baum-Welch statistics and UBM parameters. The proposed metric helps to improve the ASV performance when incorporated as a side information during system combinations. The relative improvement is considerably more when tested with short test data. This quality metric requires no additional parameters to be estimated. In our current work, we have proposed a simple scheme where the absolute differences of BW statistics and UBM parameters are used for measuring the quality. Further investigation can be conducted by adopting other dissimilarity metrics with a goal to find the optimum one. Other possible future directions include evaluating the performance in noisy conditions, compatibility test of the proposed approach with other features such as deep neural network based bottleneck features, etc.
\section*{Acknowledgment}
The authors take the opportunity to acknowledge Indian Space Research Organization (ISRO) for financing the research partially. The authors also express gratitude to Mr. Monisankha Pal and Mrs. Shefali Waldekar for technical discussions and grammatical corrections respectively.

\bibliographystyle{IEEEtran}
\bibliography{bibfile}

\begin{thebibliography}{10}
\providecommand{\url}[1]{#1}
\csname url@samestyle\endcsname
\providecommand{\newblock}{\relax}
\providecommand{\bibinfo}[2]{#2}
\providecommand{\BIBentrySTDinterwordspacing}{\spaceskip=0pt\relax}
\providecommand{\BIBentryALTinterwordstretchfactor}{4}
\providecommand{\BIBentryALTinterwordspacing}{\spaceskip=\fontdimen2\font plus
\BIBentryALTinterwordstretchfactor\fontdimen3\font minus
  \fontdimen4\font\relax}
\providecommand{\BIBforeignlanguage}[2]{{%
\expandafter\ifx\csname l@#1\endcsname\relax
\typeout{** WARNING: IEEEtran.bst: No hyphenation pattern has been}%
\typeout{** loaded for the language `#1'. Using the pattern for}%
\typeout{** the default language instead.}%
\else
\language=\csname l@#1\endcsname
\fi
#2}}
\providecommand{\BIBdecl}{\relax}
\BIBdecl

\bibitem{kinnunen2010overview}
T.~Kinnunen and H.~Li, ``An overview of text-independent speaker recognition:
  {F}rom features to supervectors,'' \emph{Speech communication}, vol.~52,
  no.~1, pp. 12--40, 2010.

\bibitem{reynolds2000speaker}
D.~A. Reynolds, T.~F. Quatieri, and R.~B. Dunn, ``Speaker verification using
  adapted {G}aussian mixture models,'' \emph{Digital signal processing},
  vol.~10, no.~1, pp. 19--41, 2000.

\bibitem{poddar2017speaker}
A.~Poddar, M.~Sahidullah, and G.~Saha, ``Speaker verification with short
  utterances: A review of challenges, trends and opportunities,'' \emph{IET
  Biometrics}, 2017.

\bibitem{arnab2015comparison}
------, ``Performance comparison of speaker recognition systems in presence of
  duration variability.'' in \emph{2015 Annual IEEE India Conference
  (INDICON)}.\hskip 1em plus 0.5em minus 0.4em\relax IEEE, 2015, pp. 1--6.

\bibitem{li2016improving}
L.~Li, D.~Wang, C.~Zhang, and T.~F. Zheng, ``Improving short utterance speaker
  recognition by modeling speech unit classes,'' \emph{IEEE/ACM Transactions on
  Audio, Speech, and Language Processing}, vol.~24, no.~6, pp. 1129--1139,
  2016.

\bibitem{kanagasundaram2011vector}
A.~Kanagasundaram, R.~Vogt, D.~B. Dean, S.~Sridharan, and M.~W. Mason,
  ``I-vector based speaker recognition on short utterances,'' in
  \emph{Proceedings of INTERSPEECH}.\hskip 1em plus 0.5em minus 0.4em\relax
  International Speech Communication Association (ISCA), 2011, pp. 2341--2344.

\bibitem{hasan2013duration}
T.~Hasan, R.~Saeidi, J.~H. Hansen, D.~van Leeuwen \emph{et~al.}, ``Duration
  mismatch compensation for i-vector based speaker recognition systems,'' in
  \emph{Acoustics, Speech and Signal Processing (ICASSP), 2013 IEEE
  International Conference on}.\hskip 1em plus 0.5em minus 0.4em\relax IEEE,
  2013, pp. 7663--7667.

\bibitem{kanagasundaram2014improving}
A.~Kanagasundaram, D.~Dean, S.~Sridharan, J.~Gonzalez-Dominguez,
  J.~Gonzalez-Rodriguez, and D.~Ramos, ``Improving short utterance i-vector
  speaker verification using utterance variance modelling and compensation
  techniques,'' \emph{Speech Communication}, vol.~59, pp. 69--82, 2014.

\bibitem{arnab2017adapt}
A.~Poddar, M.~Sahidullah, and G.~Saha, ``An adaptive i-vector extraction for
  speaker verification with short utterance,'' in \emph{in Proc. International
  Conference on Pattern Recognition and Machine Intelligence (PReMI)}.\hskip
  1em plus 0.5em minus 0.4em\relax Springer, 2017.

\bibitem{li2016feature}
W.~Li, T.~Fu, H.~You, J.~Zhu, and N.~Chen, ``Feature sparsity analysis for
  i-vector based speaker verification,'' \emph{Speech Communication}, vol.~80,
  pp. 60--70, 2016.

\bibitem{dehak2011front}
N.~Dehak, P.~Kenny, R.~Dehak, P.~Dumouchel, and P.~Ouellet, ``Front-end factor
  analysis for speaker verification,'' \emph{Audio, Speech, and Language
  Processing, IEEE Transactions on}, vol.~19, no.~4, pp. 788--798, 2011.

\bibitem{kenny2010bayesian}
P.~Kenny, ``Bayesian speaker verification with heavy-tailed priors.'' in
  \emph{Odyssey}, 2010, p.~14.

\bibitem{hasan2013crss}
T.~Hasan, S.~O. Sadjadi, G.~Liu, N.~Shokouhi, H.~Boril, and J.~H. Hansen,
  ``{CRSS} systems for 2012 {NIST} speaker recognition evaluation,'' in
  \emph{ICASSP}.\hskip 1em plus 0.5em minus 0.4em\relax IEEE, 2013, pp.
  6783--6787.

\bibitem{hautamaki2013sparse}
V.~Hautamaki, T.~Kinnunen, F.~Sedl{\'a}k, K.~A. Lee, B.~Ma, and H.~Li, ``Sparse
  classifier fusion for speaker verification,'' \emph{Audio, Speech, and
  Language Processing, IEEE Transactions on}, vol.~21, no.~8, pp. 1622--1631,
  2013.

\bibitem{brummer2013bosaris}
N.~Br{\"u}mmer and E.~De~Villiers, ``The bosaris toolkit: Theory, algorithms
  and code for surviving the new dcf,'' \emph{arXiv preprint arXiv:1304.2865},
  2013.

\bibitem{sahidullah2016local}
M.~Sahidullah and T.~Kinnunen, ``Local spectral variability features for
  speaker verification,'' \emph{Digital Signal Processing}, vol.~50, pp. 1--11,
  2016.

\bibitem{sahidullah2012design}
M.~Sahidullah and G.~Saha, ``Design, analysis and experimental evaluation of
  block based transformation in {MFCC} computation for speaker recognition,''
  \emph{Speech Communication}, vol.~54, no.~4, pp. 543--565, 2012.

\bibitem{sahidullah2013novel}
------, ``A novel windowing technique for efficient computation of {MFCC} for
  speaker recognition,'' \emph{Signal Processing Letters}, vol.~20, no.~2, pp.
  149--152, 2013.

\bibitem{arnab2017improved}
A.~Poddar, M.~Sahidullah, and G.~Saha, ``Improved i-vector extraction technique
  for speaker verification with short utterances,'' in \emph{International
  Journal of Speech Technology}.\hskip 1em plus 0.5em minus 0.4em\relax
  Springer, 2017.

\end{thebibliography}

\end{document}